# A framework for modeling interdependencies among households, businesses, and infrastructure systems; and their response to disruptions

Interdependencies & resilience of firms/households and infrastructures

*Mateusz Iwo Dubaniowski[1]\*, Hans R. Heinimann[1]*

[1]ETH Zurich, Future Resilient Systems at the Singapore-ETH Centre (SEC), which was established collaboratively between ETH Zurich and National Research Foundation (NRF) Singapore (FI 370074011), under the auspices of the NRF's Campus for Research Excellence and Technological Enterprise (CREATE) programme.

\*-corresponding author

Singapore-ETH Centre
Future Resilient Systems
1 CREATE Way
#06-01 CREATE Tower
Singapore 138602
Singapore
iwo.dubaniowski@frs.ethz.ch
+65 85911898

## Abstract

Urban systems, composed of households, businesses, and infrastructures, are continuously evolving and expanding. This has several implications because the impacts of disruptions, and the complexity and interdependence of systems, are rapidly increasing. Hence, we face a challenge in how to improve our understanding about the interdependencies among those entities, as well as their responses to disruptions.

The aims of this study were to (1) create an agent that mimics the metabolism of a business or household that obtains supplies from and provides output to infrastructure systems; (2) implement a network of agents that exchange resources, as coordinated with a price mechanism; and (3) test the responses of this prototype model to disruptions.

Our investigation resulted in the development of a business/household agent and a dynamically self-organizing mechanism of network coordination under disruption based on costs for production and transportation. Simulation experiments confirmed the feasibility of this new model for analyzing responses to disruptions. Among the nine disruption scenarios considered, in line with our expectations, the one combining the failures of infrastructure links and production processes had the most negative impact. We also identified areas for future research that focus on network topologies, mechanisms for resource allocation, and disruption generation.

# Introduction

Urban systems of infrastructure, businesses, and households are constantly expanding and evolving. An increasing amount of human activity is centered around cities, which causes populations to gravitate toward them. Consequently, the number of large cities is growing rapidly [1]. As urbanization advances, related entities become more prevalent and involved in shaping humans, businesses, infrastructure, and government interactions. This has several implications for urban systems. First, the impact of similar disruptions is heightened because a higher concentration of entities within a city is associated with a greater degree of damage that arises from disruptions of the same magnitude. "Disruption" is defined here as an unexpected, undesirable disturbance that negatively influences a system, or any, or several of its components. Second, the flow of goods and services between businesses and households is continuously increasing. These entities require access to various goods, services, and infrastructure systems if they are to survive and perform even the most basic functions. Third, recent approaches in cybernetics refer to cities as organisms that combine businesses, households, and infrastructure, all of which are becoming more involved and interdependent [2][3] due to technological advancements and the greater complexity of production processes [4][5].

This rapid urbanization and increasing dependence on urban systems present issues that are vital to inhabitants of cities. For example, the extent of the damage caused by similarly sized, simultaneous, disruptions can be larger for cities with higher densities. Furthermore, the flow of goods and services depends upon the unique profiles of households and businesses. These profiles can become more flexible and change frequently in response to factors that then lead to dynamic changes for urban metabolism. Realizing and modeling the behavior of socio-economic units within the context of infrastructure is a major challenge for societies, businesses, and governments. Therefore, it is crucial that we gain more knowledge about these interdependencies, their fluctuations, and their impact on such systems.

The objective of our study was to address the need for understanding the interdependencies among businesses, households, and infrastructure systems; as well as how they affect reactions to various disruptions. Our aims were to (1) develop an agent that represents a business or a household by mimicking the process of transforming a set of supplied goods and services into a set of output goods and services, (2) devise a network of these agents that is dynamically self-organizing under a disruption based on varying costs of resources throughout the network, and (3) to introduce disruptions to these systems and then investigate how that might influence performance. Follow-up experiments allowed us to evaluate the feasibility of our model for assessing disruption-related changes in performance by networks. Our ultimate goal was to improve the ability of planners and managers to prepare for system failures by recognizing which types of disruptions are the most threatening and could have the most severe impact on a system of interest.

The resilience of infrastructure and economic communities is quickly becoming a vital feature of urban systems. Historically, risk management was the primary approach taken for predicting and dealing with disruptions in such settings. However, new scenarios have arisen for which the concept of risk management is less feasible. For increasingly interdependent systems, we cannot predict disruptions and their impacts as accurately as was done previously. Conventional approaches to risk management often overlook unexpected, system-wide threats and are not concerned with system recovery. In fact, employing risk-based protocols can be costly and ineffective against sudden disruption-generating events [6][7][8]. Such disruptions can take various forms, from carefully planned attacks to natural disasters. In highly interconnected societies, even a small disruption to one part of one system can propagate and generate extremely negative consequences for other systems [9]. As a result, shocks to a system become more frequent and more detrimental to households and businesses [10]. Hence, the concept of resilience, originating in a system's resistance to and recovery from severe or unexpected disruptions, is now the defining characteristic of urban systems and is also being introduced and applied to other fields of research [5]. National Academy of Sciences (NAS) defines resilience as "the ability to anticipate, prepare for, and adapt to changing conditions and withstand, respond to, and recover rapidly from disruptions." The defining features of resilience are the focus on recovery from and adaptation to disruptions. These are extending the concept of risk, which emphasizes robustness of the system i.e. being able to withstand a disruption as is. Furthermore, the risk assessment and quantification are usually done with respect to particular events that affect system and its components, while resilience describes system's ability to survive any unlikely disruptive event due to the system's inherent structure and activities [11]. It is also important to note that recovery part of this definition includes the dynamic nature of resilience, where the system responds to an event in a way that preserves its critical functionality within reasonable time. Thus, providing a speedy recovery is a crucial feature of a resilient system [12]. This is in contrast with risk, where the temporal nature of recovery over time is not necessarily captured.

Contingency plans and response scenarios can make systems more resilient. However, having a better understanding of a disrupted system can aid in estimating disruption related costs. The process of recovering can also be modeled and illuminated by examining how a system might adapt. To model these unexpected disruptions, one must consider the following factors: (1) behavior of businesses and households that mimics their interactions with infrastructure systems, and (2) dynamically variable interdependencies among those interacting components.

Several streams of research have described the recovery and response of infrastructure to disruptions [13]. Those efforts include a focus on modeling resilience of individual systems for electrical power [14][15], water supplies [16], or transportation [17][18][19]. However, they also include attempts to model resilience of urban infrastructures in general [20][21]. Another stream [22][23] is devoted to simulating various interdependencies, including agent-based models [24][25][26] and system-of-systems (SoS) approaches [27]. Although input–output models can be used to describe interdependencies among infrastructure systems [28][29][30][31][32], none of these streams accounts for the interactions of multiple infrastructures with business and household agents, which would help provide insight into differences in vulnerabilities between population groups, an opportunity overlooked in regional models [33]. These include variations in the impact of and response to disruptions based on income level, health status, or type of business. For example, households that contain disabled and low-income members respond differently to disruptions when compared with households made up of fully healthy persons or those with high incomes. Similarly, retail stores are affected differently to a restaurant when a power outage occurs.

Interdependent networks of infrastructure as well as individual infrastructure systems have been analyzed within the context of graph theory. The connectiveness of such graphs and other topological properties of these graphs have been examined. These measures can be used to understand vulnerabilities within infrastructure system networks such as electric grids [34][35], or urban street networks [36]. However, these mathematical methods do not consider the costs of producing resources, or any interactions among various interdependent infrastructure networks, businesses, and households, as well as the dynamic nature of disruptions.

In looking at the internal workings of businesses and households, the input–output model developed by Leontief [37] has long been used to illustrate their behavior [38][39]. However, this type of model has not been applied to simulate such interactions within an SoS setting. Instead, other models of household and business behavior have incorporated more insight into their internal social organizations, rather than relying upon the effects of external production, as is the case in the input–output model [40][41]. In modeling disruptions to systems, researchers have examined the impact on supply-chain networks and developed useful frameworks [42][43][44]. Disruptions to the infrastructure have been analyzed with water supplies [45] and healthcare services [46]. A World Bank report [32], combines the input-output model and network criticality analysis to estimate the impact of disruptions to the transportation system in Tanzania. This approach, however, does not include any other infrastructure systems besides transportation network and focuses primarily on supply-chains within the country rather than interdependencies between infrastructures. Consequently, the model outlined by Colon et al. has limited application to modeling impact of disruptions on multiple infrastructure networks.

## Model specification

We developed an agent-based network that formed an SoS for the flow of goods and services among all of those agents. This flow was meant to represent the relationships and interdependencies among different actors within a society. The system-of-systems reflects the multitude of different independent infrastructure systems included in one simulation that involves businesses and households, i.e., socio-economic systems. Examples of an infrastructure system would be power grids, water or gas supplies, transportation, or telecommunications.

Our specifications called for a conceptual model that could simulate the flows of resources throughout a network of socio-economic agents and infrastructure systems. The new framework and its components consisted of agents and internal working mechanisms of the network. It also included a means by which one could generate network disruptions and identify the factors possibly responsible. Furthermore, the framework featured a coordination system that managed resource flows among the various network agents.

In our model all resources including infrastructure, production, business and household resources are introduced or produced by agents, with agents being joined together to form networks. Agents are linked together under this model with infrastructure links that provide means for transfer of resources between the agents. This results in interdependencies in the system being represented in two ways: (1) through infrastructure links; (2) through matrices of technical coefficients within agents.

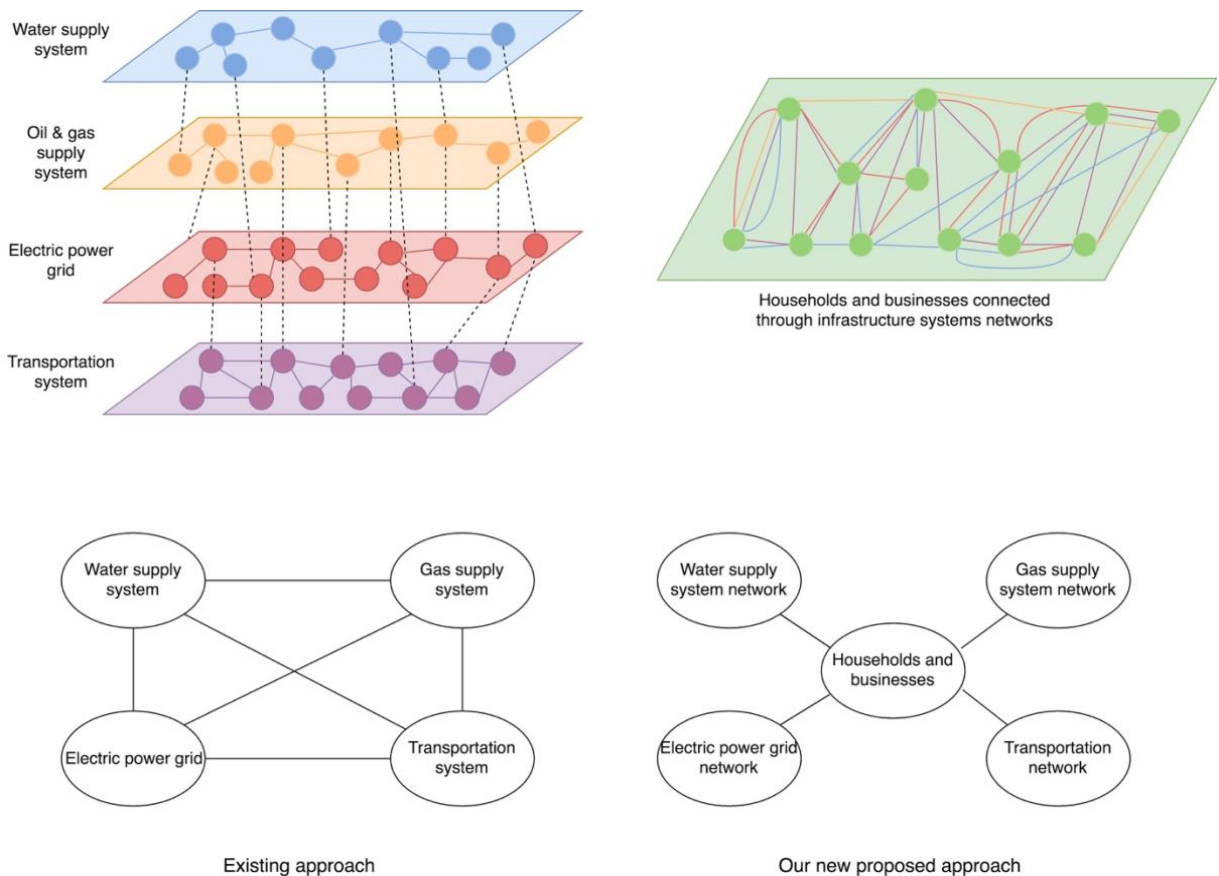

Figure 1: Comparison of existing approach for modelling infrastructure systems and their resilience versus new approach. In latter, households and businesses are included to indicate interdependencies between systems. Infrastructures are represented through network links and their corresponding infrastructure business agents.

The comparison of our proposed approach with an existing model is shown in FIGURE 1. For our model, interdependencies among infrastructure systems were manifested through households and businesses, with the latter also providing infrastructure. Thus, infrastructure is represented by both the infrastructure network and the business agents that ensure production of the resource delivered through the infrastructure network. In the existing approach, interdependencies were defined between each pair of infrastructure systems, which themselves operated independently. In our model, the dependencies among socio-economic units included combinations of resources for various infrastructure systems that would then be used to produce other resources.

For example, to provide water, the supply company represented by an agent required electrical energy, petrol, gas, transportation and maintenance equipment (capital goods), and consumer goods that their employees might consume at work. The producer distributed the water through a supply network connected to that company agent. If, due to some adverse conditions, the energy supply to that company became more expensive, the cost to provide water would increase. This effect was then propagated throughout the entire water network connected to the supplier company agent, thereby influencing all of the following agents that were dependent on the water supply from this source. These included the agent for the power company, which needed water to generate electricity. Increasing the cost of water sent to the power supply agent increased the price of electricity even further, again leading to a propagation of that effect through the network to other household and business agents. Therefore, we observed a feedback loop representing the interdependency of these two systems. Moreover, the increasing costs seen by both households and other businesses that received those essential sources resulted in higher prices for the resources produced or consumed by those household and business agents.

## Conceptual framework

Our network of agents corresponded to households or businesses that exchanged goods and services through links that represented flows through an infrastructure system. Those agents could produce as well as consume the resources. This production process entailed a set of inputs being transformed into a set of outputs. The steps that could be performed were defined uniquely for each agent, who then took inputs from the network and converted them into outputs that would satisfy the demand for goods and services by another agent. This transfer and exchange from one agent to the next was accomplished over a network of infrastructure systems that corresponded to physical and socio-technical links between agents. They included roads, telecommunication lines, pipelines, power grids, and similar components. The infrastructure links corresponded to the edges of the agent network. Each edge was associated with a cost vector that specified transportation costs per unit for each resource over that edge.

The disruptions introduced here were stochastically modeled as discrete events. We used different methods depending upon whether we wanted to simulate a disruption to a production process, an infrastructure system, or the external demand of the entire system. Agents were connected through infrastructure links that coordinated the flow of goods and services with a pricing mechanism that governed how and where resources moved, and, effectively, where production occurred. The aim was to minimize the overall costs of satisfying total demand from the system. Simulation of the model was dynamic and ran in discrete timesteps that constantly adjusted to any possible disruptions introduced to the system.

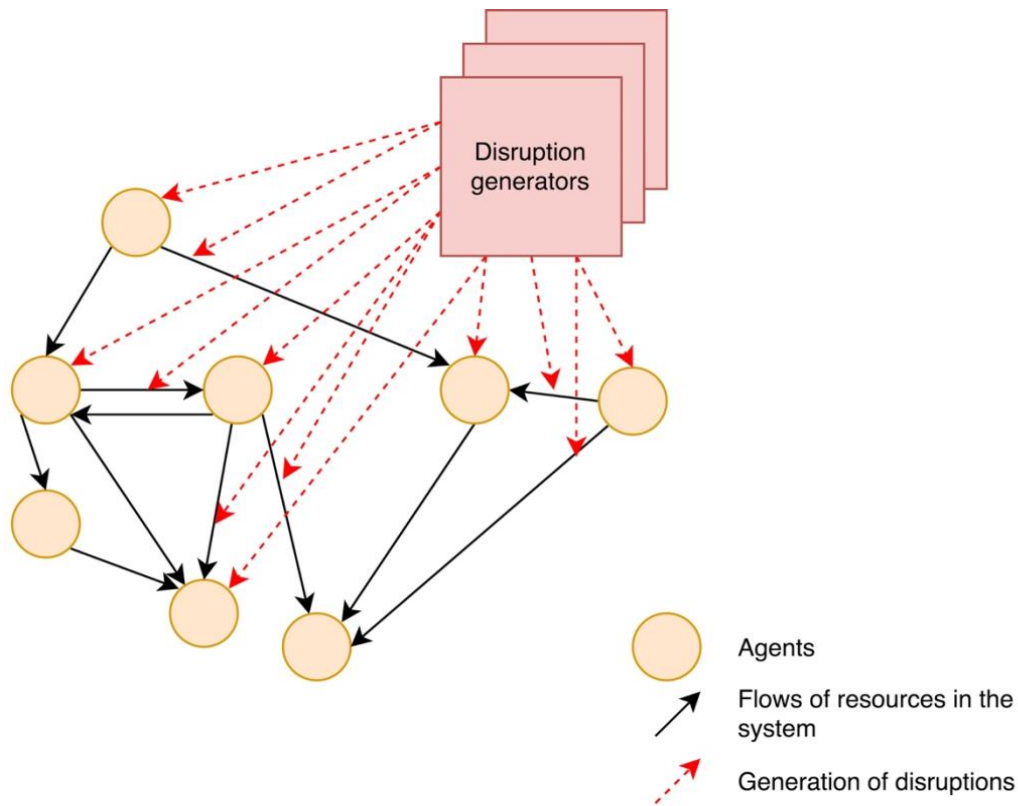

Figure 2: Conceptual framework of business/household agents interacting with infrastructure systems while facing disruptions. Network of agents exchange goods and services through links. Both agents and links could be interrupted by disruption generators.

The conceptual framework is shown in FIGURE 2. Agents exchanged goods and services through edges that were responsible for resource flow between agents, based on a pricing mechanism to minimize costs. Those edges corresponded to network components, e.g., power grids, transportation, water supply, and telecommunications connections. Disruptions were generated and introduced at various points within the SoS. Emergent behavior was observed under various disruptions, during which agents could perform different functions based on their unique advantages relative to other agents post-disruption. Together, the network of agents, coordination mechanism, and disruption generator formed a system that dynamically reorganized and coordinated the network under a disruption. The SoS was able to model interdependent responses to disruptions that individual systems, if simulated separately, might have failed to capture.

Our new model assumed that individual socio-economic agents and individual infrastructure system links behaved linearly at each timestep in the simulation. This means that resources produced by each agent or transported through a link depended on the inputs to the agent or the link, since the behavior of an agent followed an essentially linear input-output model. This served as a fair approximation of the real world where, in the absence of any major disruptions, infrastructure systems and businesses would individually operate linearly at sufficiently short time windows. Even though our model was linear in the way that individual socio-economic agents, and infrastructure system links behaved, the links between these components forming a network could form feedback loops or flows that could result in the overall system's non-linear behavior.

Similarly, the disruption generators were intended to be non-deterministic and to reflect the dynamic nature of the entire SoS under a disruption or due to passage of time. Those generators represented dynamic changes to both infrastructure system links and to the internal workings of the socio-economic agents. Their disruptions were non-deterministic, periodic, or one-time events; recoverable or not; and could have negative or even positive impacts on the components of the system that they disrupted. Despite the static nature of socio-economic agents and the coordination mechanism itself, the overall model was dynamic due to the introduction of disruptions progressing and cascading over time.

The linearity within short timeframes is a fair assumption. Approximating production with a linear model is adequate since the demands and production of a society in a short time window to a large extent exhibits linear behavior and can be described with a linear model such as input-output model. Throughout the day there are several periods, which are characterized by a particular linear relationship such as in the morning, where commuters travel to work, and companies do not operate yet, and throughout the day where most businesses operate at full capacity. These periods can be approximated with a linear model each at a relevant time granularity. At the time granularity, where the time window we propose to be 15 minutes, the linearity assumption is valid. The production processes within such time window can be approximated and characterized by a linear model for most infrastructure systems and businesses. Nevertheless, the time granularity of the model can be adjusted accordingly depending on the proposed socioeconomic systems to be modeled.

Similarly, at each timestep the system is static as no system dynamics mechanisms take place within that single timestep. The input and outputs of the model for each timestep are fixed. The introduction of disruption generators introduces dynamic behavior into the system by changing the system's components between subsequent timesteps. The interdependencies between infrastructure systems and businesses at each timestep are static, while introduction of disruptions, which alter the system's properties between individual timesteps, results in the overall system's dynamic behavior. Thus, the system with disruption generators introduced is dynamic. The size of the individual timestep, i.e. time granularity, needs to be carefully adjusted to ensure that the dynamic properties of the real-world systems are adequately represented in the model [47]. There exists also a trade-off between time granularity and speed of the simulation, which needs to be considered when applying the model to a particular scenario.

### Agent specification
We defined a socio-economic agent as a single unit – business or household – that manifested economic activity in terms of goods and services produced or consumed by that agent. As such, it was the smallest unit in the SoS model.

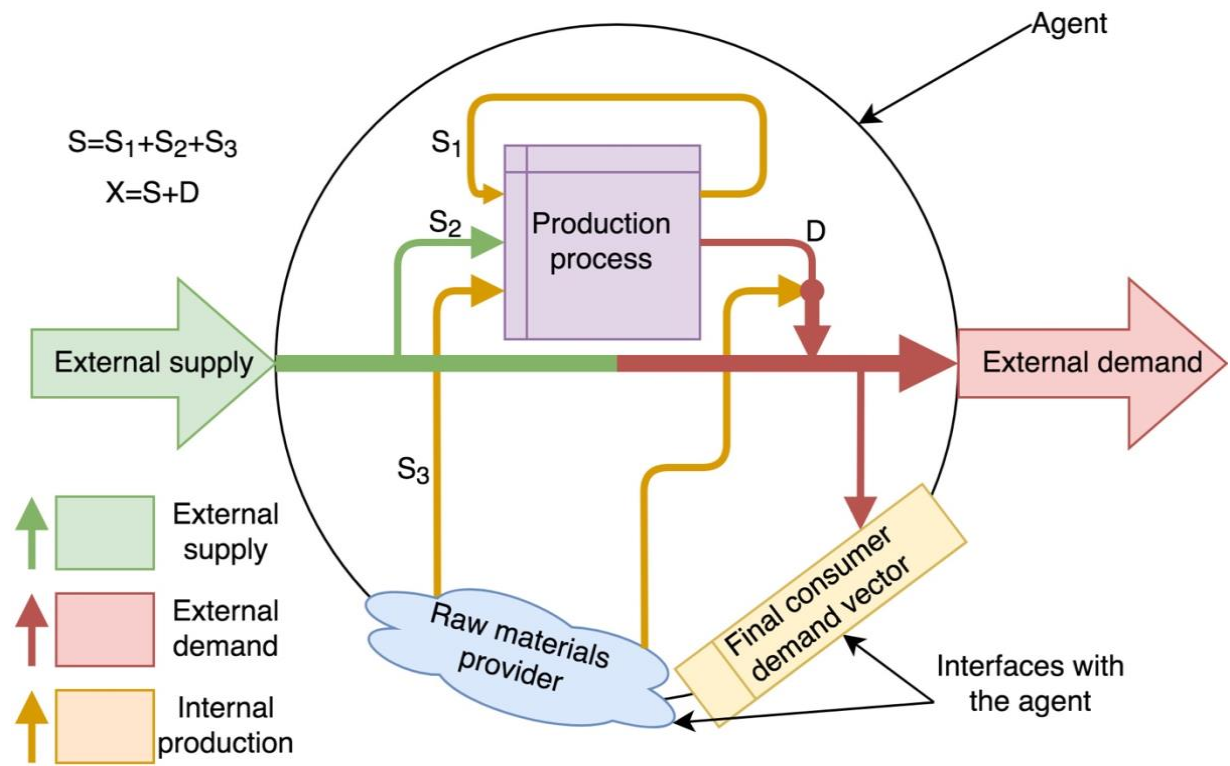

Figure 3: Functionality of household/business agent (signified by circle on diagram) that receives input resources from preceding agents (external supply – $S_2$) and performs production processes on these resources to satisfy demand by next agents (external demand - D). The internal production for self-consumption is also visible ($S_1$). Agent can also introduce raw materials ($S_3$) and be the final consumer through interfaces with provider and demand vector, or can also serve as through-transportation link, conveying resources to subsequent agents without performing production processes.

Reflecting real-world conditions, an agent took in a set of inputs (external supply) and produced a set of outputs (external demand) through a prescribed process. The agent's behavior was primarily defined by that external demand together with the final consumer-demand vector. The production process was modeled by a technology matrix [37][48] to determine what quantities of which resources were needed to produce a single unit of another resource. Thus, it served as the "brain" of an agent. This process, as well as the agent, could also supply resources to itself (FIGURE 3). For example, an intermediate step might require a resource produced by the agent that was then used in the final step by the same agent. Some agents also had the capacity to introduce raw materials through a provider component. The costs associated with supplying each resource could differ among agents according to different price levels. Finally, the agent was connected to a final consumer-demand vector that represented the sink, or ultimate consuming capacity, of the system. This vector was used to show which resources an agent consumed without producing anything in return. Such consumer agents simulated the overall, aggregate external demand of the entire SoS. The difference between external demand and final demand from the agent's point of view was irrelevant; both were combined to form the total demand from the agent. However, in terms of the entire network, their purposes differed. Similarly, the provider of raw materials and the external supply were uniform from the agent's perspective. However, from the network's perspective, they corresponded to different functions and properties, as clearly demonstrated in FIGURE 3.

## Production process

The production process was represented according to the technology matrix originally suggested by Leontief [37] and elaborated upon by Lin and Polenske [38] and Albino, Izzo, and Kühtz [49]. This matrix described the quantities of each resource needed to produce a single unit of another resource. Those inputs were transformed into outputs based on the matrix entries (example shown in Table 1). An agent could perform a production process by utilizing its technology matrix, or conduct other tasks to satisfy its external demand. The following formulae were used with this matrix to determine the cost and amount of each necessary resource:

$$X = AX + D$$
$$(I - A)X = D$$
$$X = (I - A)^{-1}D$$
$$S = X - D$$

Where, $S$ represents the supply (all inputs) to a production process, $X$ is the production vector (i.e., the sum of all inputs and external outputs), $A$ is the matrix of technical coefficients, $(I - A)$ is the technology matrix [48], $(I - A)^{-1}$ is calculated as Leontief's inverse technology matrix [48], and $D$ represents the external demand (outputs) of the process. The above variables are annotated and linked to corresponding flows on FIGURE 3.

Solving these equations for the process used by each agent, as well as the specified level of external demand for each agent, allowed us to identify which inputs and resources must be supplied to each agent to satisfy production demands at the lowest cost. For agents not connected to the final consumer-demand vector, the technology matrix was the most important component for an individual agent because it could be manipulated or replaced throughout the simulation either to show dynamic changes in production processes or demonstrate processes undergoing periodic variations or disruptions.

## Disruptions to the system

The model described in this paper was designed to help predict the effects of disruptions on a system, as well as to predict its recovery afterward. As such, disruption generation was integral to the model. Recovery and response differed according to how and where a disruption occurred, such as when introduced to production processes or infrastructure system links. Moreover, demands from the system could change drastically, thus initiating a disruption. Real-world disruptions translate into corresponding disruption representations in our model (FIGURE 4).

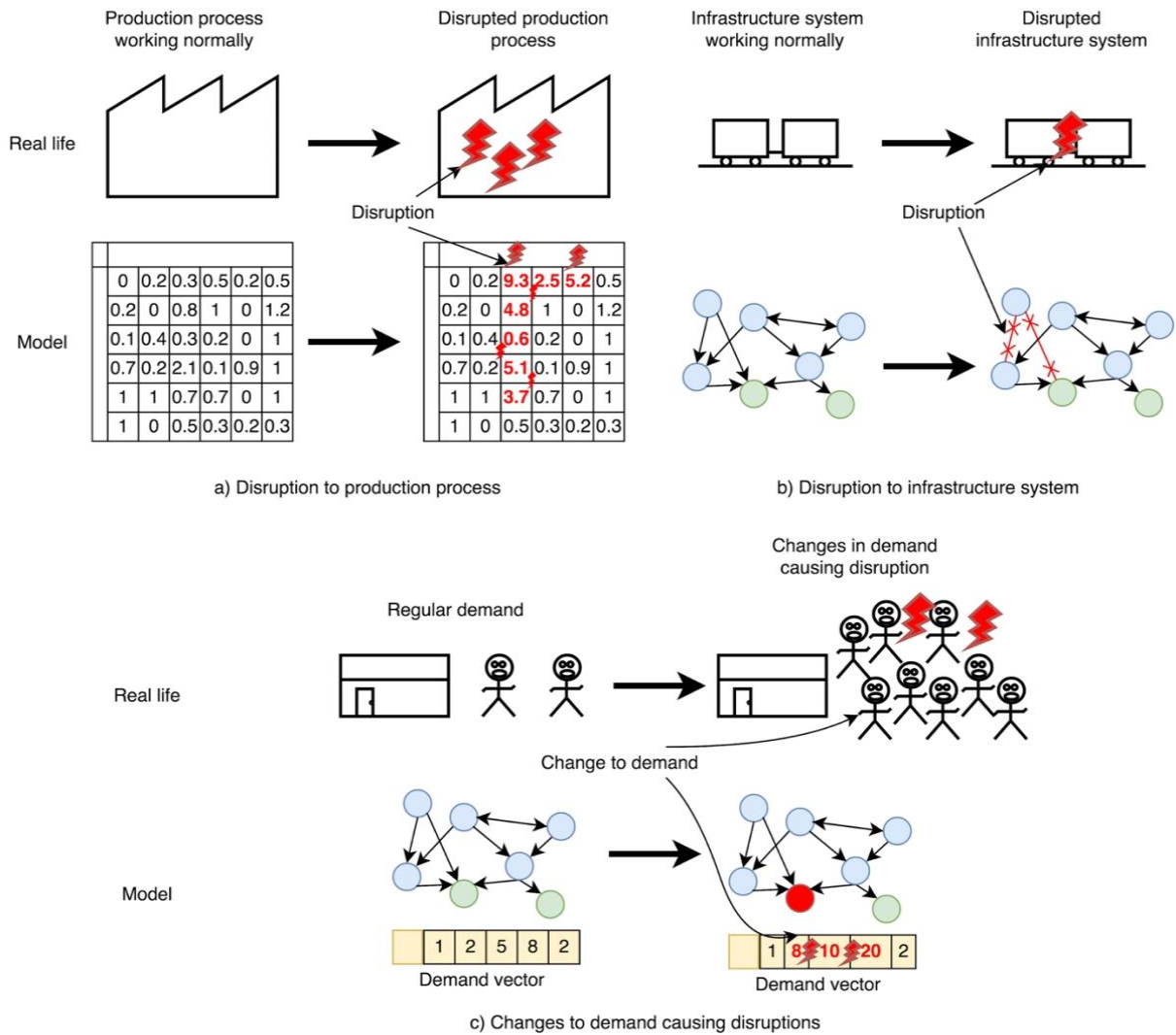

**Figure 4: Three types of disruptions with their correspondence in real world: a) disruption to production process, b) disruption to infrastructure network links, c) disruption due to change in demand.**

A disruption in our model was designed to be non-deterministic and random, thereby influencing the workings of the system. In particular, it could affect the production process(es) of an agent(s), the infrastructure system network, demand from the system, or any combination of these. It was generated according to certain governing criteria. In the real world, disruptions correspond to natural or man-made hazards that occur within a system. They can include earthquakes, hurricanes, terrorist attacks, or cascading equipment malfunctions. Here, such disruptions followed a stochastic process and were introduced as a discrete event that might then either progress or retract, depending upon the intended nature of the disruption. The system responded to the onset of a disruption by adjusting its internal workings, thus resisting the challenge. Resources were then redirected in accordance with the coordination mechanism due to changes in production, transportation, or demand caused by the disruption. A resilient system was considered one that could sustain a wide range of disruptions at a reasonable cost either through its robustness or adaptation to the disruption. In contrast, a system lacking resilience would crash or incur high costs due to difficulties in adapting to the new situation. One purpose for our model was to assess this degree of resilience. Systems with high resilience would incur small increases of supply curve for a wide range of disruptions, while less resilience would suffer higher increase. This process would allow us to compare two or more systems in terms of their resilience understood both as robustness and as ability to adapt to a disruption. The ones with the highest increase of supply costs on average being the least resilient.

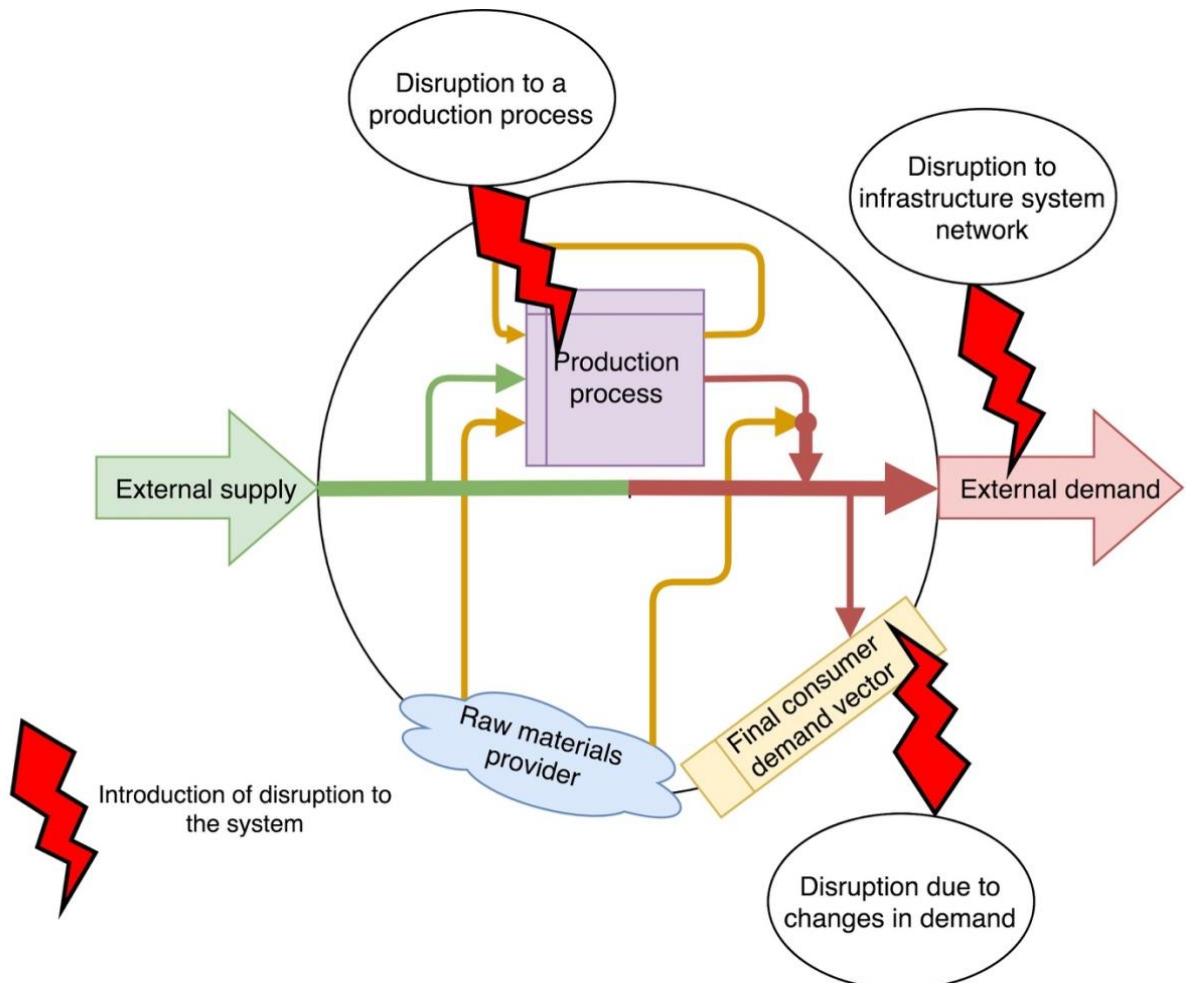

**Figure 5: Types of disruptions: agent within production process (1), infrastructure links between agents (2), or unexpected (extreme) changes in final consumer demand (3).**

As illustrated in FIGURE 4 and FIGURE 5, disruptions due to failures in production processes could include broken machinery (business or manufacturer) or a faulty water faucet (household). This required changes in the relevant fields of the technology matrix. Consequently, the need might increase for some resources that were necessary to produce another resource (here depicted as a change in a column of the technology matrix). Alternatively, this could mean an enhanced demand for the same resource regardless of what was being produced (i.e., a change in a row of the matrix).

Introducing a disruption that would vary the production process's technology matrix required a thorough knowledge of that process itself. For example, a particular type of disruption might develop or certain fields within the matrix for each agent might be coupled and then be involved in the same kind of disruption.

Disruptions could also affect various aspects of infrastructure, e.g., cuts to power lines, or failures in telecommunications links or in the process of delivering resources. These possibilities were represented in our system by removing or decreasing the capacity of links between agents, or by increasing the cost of transporting certain resources over a particular link. Such failures could induce a rise in the cost of resource distribution or, in extreme cases, a complete failure to deliver goods and services demanded by an agent. In particular, any failure introduced to edges, i.e., links between agents, could result in higher costs or a total inability to deliver specific goods unless that link had not been utilized prior to the disruption.

Dynamic changes in agent demands, or the sudden need for a large quantity of one resource type, could also disrupt a system, thereby increasing the cost of that resource or even causing a system failure. Before applying the generator for any of these types of disruptions, the modeler had to understand how those infrastructure systems were linked together, which of those links were likely to fail simultaneously, how patterns of demand were shaped, and which resources an end-user might want to consume, and are the most critical.

Disruptions in the model were introduced through a random generator that followed specific principles depending upon the type of event it was mimicking. Various generators could model different events and affect different components of the system. Those generators were non-deterministic and followed processes meant to add randomness and dynamic behavior to the model. They could also include any possible description of the recovery process, which then alleviated the impact of that disruption based on prescribed principles that followed a stochastic process.

## Coordination mechanism based on pricing

The coordination and allocation of resources between agents was based on resource prices, which were computed according to the costs to produce ($C_P$) and transport ($C_T$) both intermediary and raw materials that were inserted by providers into the system through agent interfaces. Each transportation link within an infrastructure system had a cost associated with each resource that could be moved via that link. For our model, an agent selected individual suppliers in a way that reduced the cost of producing each resource. In particular, each agent autonomously picked the cheapest available supplier to supply resources to the agent i.e. the agent picked resource suppliers to obtain each required resource from a transportation link that had the lowest cost of providing this resource to the agent. In turn, the process of each agent autonomously attempting to minimize the cost of resources produced by that agent resulted in minimizing the overall cost of satisfying the final consumption demands of the system. The costs incurred for obtaining various quantities of resources by different agents were calculated as shown below:

$$Production\ cost = C_P + C_T$$

The algorithm used to calculate the production cost is based on technology matrix of the agent and costs of all the required resources at the agent. The technology matrix determines the share of costs of each input resource in the cost of the output resource. If the incoming edges to the agent have limited capacity and cannot satisfy the desired input requirements, the output is limited and assigned to outgoing edges based on a priority list of the outgoing edges. This approach resembles what often happens in infrastructure systems, when administrative or arbitrary prioritization is enforced in case of insufficient supply.

In particular, the production cost/resource allocation algorithm follows the following pattern:

1. Calculate the desired inputs to the agent based on the demanded outputs from the agent.
2. If the calculated input is above the available input, recalculate the outputs with limits placed on the input.
3. Calculate the production cost of resources based on marginal price of inputs and the technology matrix. Pass this to the outgoing edges.
4. Pass the calculated demanded inputs onto the cheapest available incoming edges.
5. If output was limited by inputs, update the available capacity and price at the outgoing edges following the order of the priority list of the outgoing edges.
6. Perform steps 1-5 for all agents.
7. Perform steps 1-6 until system is stable i.e. all supply chain paths for all agents are determined.

When implemented, the model derived the distribution of production costs among agents, which could then be aggregated and averaged across all of the resources produced to render a supply curve for the entire system. "Demand" represented the willingness of a consumer to pay for a given resource. A demand curve was obtained for the entire system by aggregating the willingness-to-pay data across all consumers and all resources, which then indicated the overall distribution of consumers' willingness. By overlaying these two curves (FIGURE 6), we can identify the point of equilibrium for producing resources in terms of quantity and price. Our proposed model then allowed us to shape the supply curve by varying the final consumer demands in the simulation. By manipulating the quantity of final resources produced by each consumer agent we shaped the distribution of production costs across different final consumers, thus obtaining the supply curve for the system. However, our model did not enable the derivation of the demand curve. The jumps in the production cost curve, on FIGURE 6, correspond to the fact that unit cost of resources for a certain range of quantities is often the same at a given agent. The cost only varies between the agents or when quantity consumed significantly increases. E.g. a unit cost of electricity at a household is in most cases the same no matter how much is consumed, however, it can differ between households as transportation costs or suppliers are different. Hence, while everything else stays constant, a household that consumes a certain amount of resources would have constant unit marginal cost for these no matter how much it consumes unless the demand pattern of the household changes drastically.

Under our model the allocation of resources was based on costs, so that when agent faced several suppliers with differing costs, the cheapest supplier was selected. Our model assumed that the agent can always choose the lowest cost, and can switch suppliers easily and freely during the simulation. These limitations might not necessarily be true in a real-world scenario; however, they present good approximation and are of importance in networks with greater reconfigurability where switch of suppliers can happen rapidly and quickly. In less reconfigurable networks, such as water supply or power grids, there is often only one supplier for a given resource, and so this does not pose a significant limitation. When output from an agent is larger than the amount of resources that the agent can provide, the allocation of resources is done based on priority list of edges. This is another limitation that resembles infrastructure systems, where oftentimes in the case of limited supply, a prioritization of supply lines is performed. Finally, our model assumes that bottlenecks occur at transportation lines, which have certain maximum capacities. On the other hand, the agents have unlimited production capacity limited only by the capacity of input and output links.

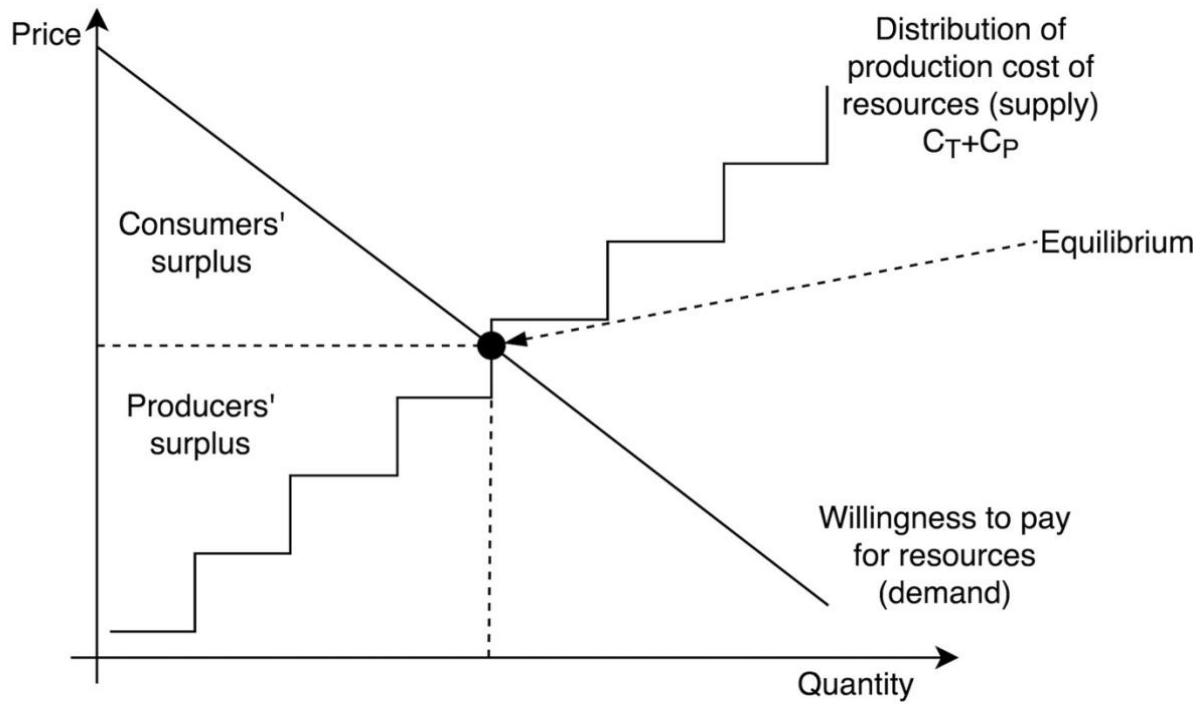

Figure 6: Demand curve represents distribution of willingness by consumers to pay for resources. Supply curve represents assignment of production costs across consumer agents. Production costs correspond to transportation cost for intermediate resources ($C_T$) + cost to obtain intermediate resources ($C_P$). In proposed model, supply curve can be derived, and overlaid with demand curve to identify equilibrium of system in terms of quantity and price of resources.

## Implementation of the system

The simulation was developed in Python 3.5.0, and was run and evaluated under Anaconda 2.4.0 Python distribution [50] on the Mac OS Yosemite 10.10.5 operating system. This network simulation and implementation was supported by the *igraph* library, version 0.7.1 [51]. Linear algebra operations were performed using the *numpy* library, version 1.10.1 [52]. The networks were stored in the Graph Modelling Language format. Results were displayed with an interfacing web page developed in JavaScript and HTML, using the *CanvasJS* library [53] to visualize the flows of goods and services and changes in costs in the form of dynamic bar charts.

System development followed the complementary modules presented above. For example, the pricing mechanism selected incoming edges with the lowest costs for resource delivery while the production process was solved using linear algebra libraries to define the required inputs to the agent based on required outputs and costs associated with resource production. Those costs were then dispersed to the outgoing edges. For an individual agent, interfaces were arranged between the providers of raw materials and the final consumer-demand vector, all of which could be easily adjusted. Disruption generators followed a stochastic process that changed technology matrices, edge parameters, and final consumer demands accordingly to simulate those interruptions.

This system was evaluated under different network topologies and various agent parameters with several disruption simulators. The topologies corresponded to infrastructure systems topologies that are found in real world. The topologies used for evaluation were devised with the use of Erdos-Renyi algorithm for network generation. We found that topologies could affect the results of the simulation as the dispersion of results throughout the system depended on topologies. Hence, careful selection and representation of topologies in the model is important to obtain meaningful simulation results for a combination of different infrastructure systems.

The system was initially tested with an extremely small network size of only three nodes that exchanged only three types of resources and ignored the capacity of the links. This sample system is shown on FIGURE 7, with clearly marked matrices of technical coefficients, transport cost vectors, a provider of raw materials, and a final consumer demand vector. Under such a scenario, the system distributed resources as expected, i.e., in line with the price mechanism. There, resource R3 was to be produced by agent A3, and R2 by A2. This was exactly how the system did perform. Resource R1 was introduced as raw material to A1. Similarly, resource R2 could have been introduced as raw material to A1, but it was more efficiently produced by A2. Because the system reflected this as expected, it passed this simple validation. The response to basic disruptions by such a small system was also analyzed after introducing a disruption scenario (full break within link T2 and an increase in matrix values for A1 in FIGURE 7). It also involved randomly generated disruptions. The system again performed as expected, decreasing in performance when disruptions were introduced correspondingly to the perceived severity of disruption, and recovering when disruptions were subsequently removed. Moreover, the system could be easily integrated with various disruption generators. Likewise, the providers of external raw materials and models of final consumer demand were easily integrated into the system with a clear interface.

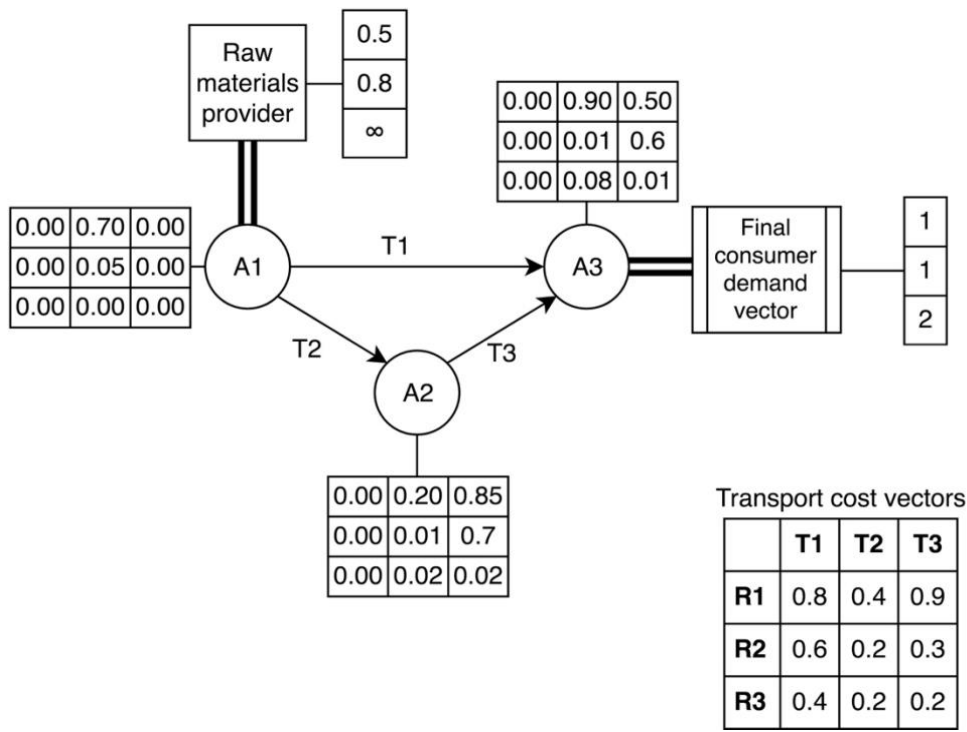

Figure 7: Initial validation of model. Sample system consists of 3 nodes connected with 3 links transferring 3 types of resources. Transportation cost vectors are shown in small tables on the Figure. Matrices of technical coefficients appear next to nodes. One agent is connected to final consumer demand; another, to provider of raw materials.

For slightly larger systems, such as the system under investigation, the outcome could not always be so easily identified analytically. Consequently, we conducted a limited validation that tested whether the allocation of resources and the path that these resources followed throughout the network was indeed optimal. For this, the system passed the validation. However, for larger systems and random, non-deterministic components and disruptions, it would have been impossible to perform conclusive testing because, except for extreme cases (e.g., intentional breakdown of the system) where SoS-wide expectations and measures of performance would be clear, we could only broadly define the expected outcome here. The difficulty of validating and testing an SoS might be attributable to trouble associated with assessing expectations, and the metrics of those expectations, for the system. Even when they can be defined, such testing can be difficult in a conventional sense [54]. Likewise, in this case, the theoretical outcome of disruptions was not known beforehand because the very purpose of the simulation was to assess that outcome. If that result had been easy to obtain prior to the disruption, then the SoS simulation would have been redundant. Hence, validating the system under larger test-case scenarios was performed qualitatively, based on whether a more-complex system would respond to disruptions in a similar fashion to the very small scenarios of just three nodes.

## Simulation experiment

We designed and ran a 3x3 factorial simulation experiment to explore the validity of our model. The consequences of predefined disruptions were assessed using cost as the main measure of performance.

## System under investigation

As proof of concept, we developed a small-scale simulation of this proposed model applied to an abstract geographic area that represented a single block of streets in an urban setting (FIGURE 8). This area incorporated a network of household and business agents connected by links through infrastructure systems. Those links supported the transportation of resources between agents that then performed production processes. By assigning parameters to each component of the system, we defined a technology matrix corresponding to a production process for each agent. This network also included transportation cost vectors for each link. After establishing the raw-material providers and final consumer-demand vectors, we generated disruption scenarios to simulate how system performance might change under such circumstances.

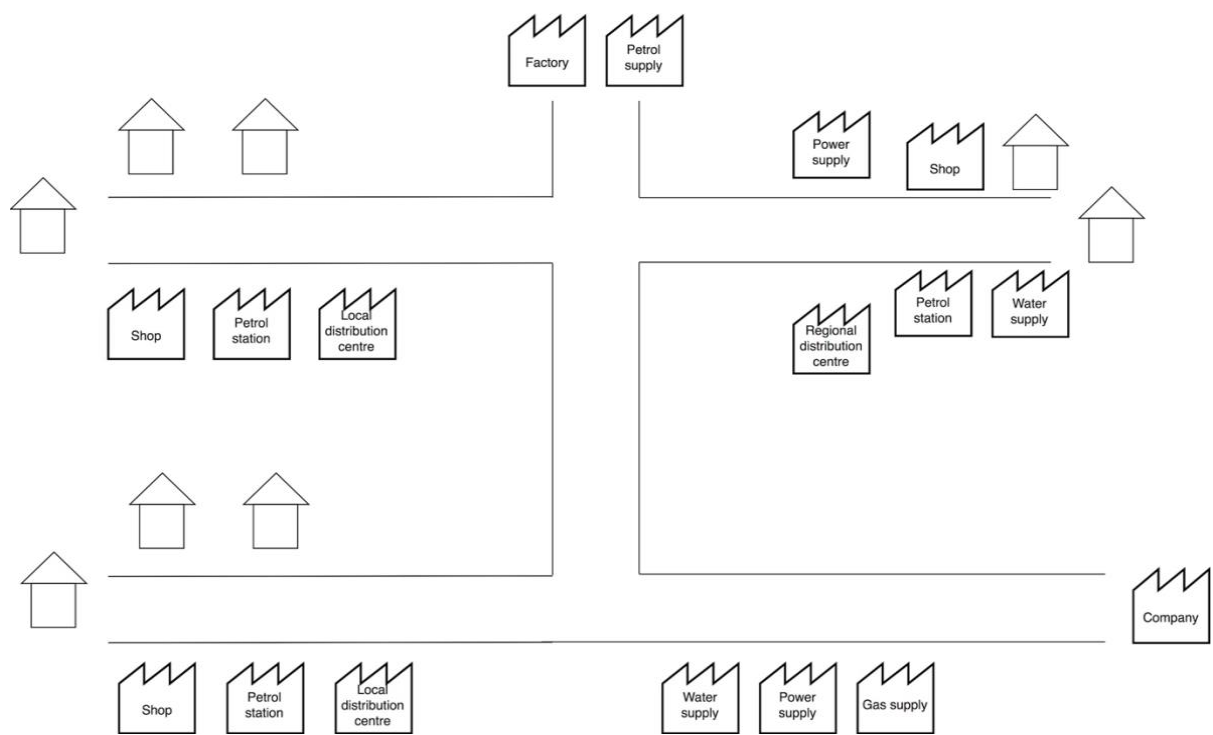

**Figure 8: Case study for simulation experiments involving single test block within urban geographical area comprising interconnected households, businesses, road network, and suppliers of electrical power, water, gas, and petrol.**

The network comprised 14 agents, i.e., eight for households and six for businesses, that might be found within one block in a certain urban neighborhood (FIGURE 9). The role of each agent was to represent particular type or types of businesses or household in the particular location. The business agents either were individual companies or else they aggregated several, multi-functional businesses in the same location, the latter type serving as a single simulation agent. For each location the household agents represented units with unique population characteristics, differentiated by income level. Agents exchanged resources with each other through 16 edges that stood for the infrastructure links between those households and businesses. The system simulated the exchange, transportation, and production of six different resources: electrical power, water, gas, petrol, capital goods, and consumer goods/services. These resources were selected as the most important to urban areas and covering as many infrastructures as possible to ensure the widest coverage of interdependencies between infrastructures and socioeconomic units. Electrical power is vital to almost all production processes and to survival of households that need electricity to operate technology and to perform all tasks. Water is likewise crucial to survival of cities, where cities cannot last longer than few days when water supply is restricted. Similarly, gas and petrol are crucial to transportation and many business processes, as well as heating of households. Consumer and capital goods represent the products of businesses that are consumed primarily by households or businesses respectively. Consumer and capital goods are at the heart of modeling interdependencies of businesses, households and infrastructure systems.

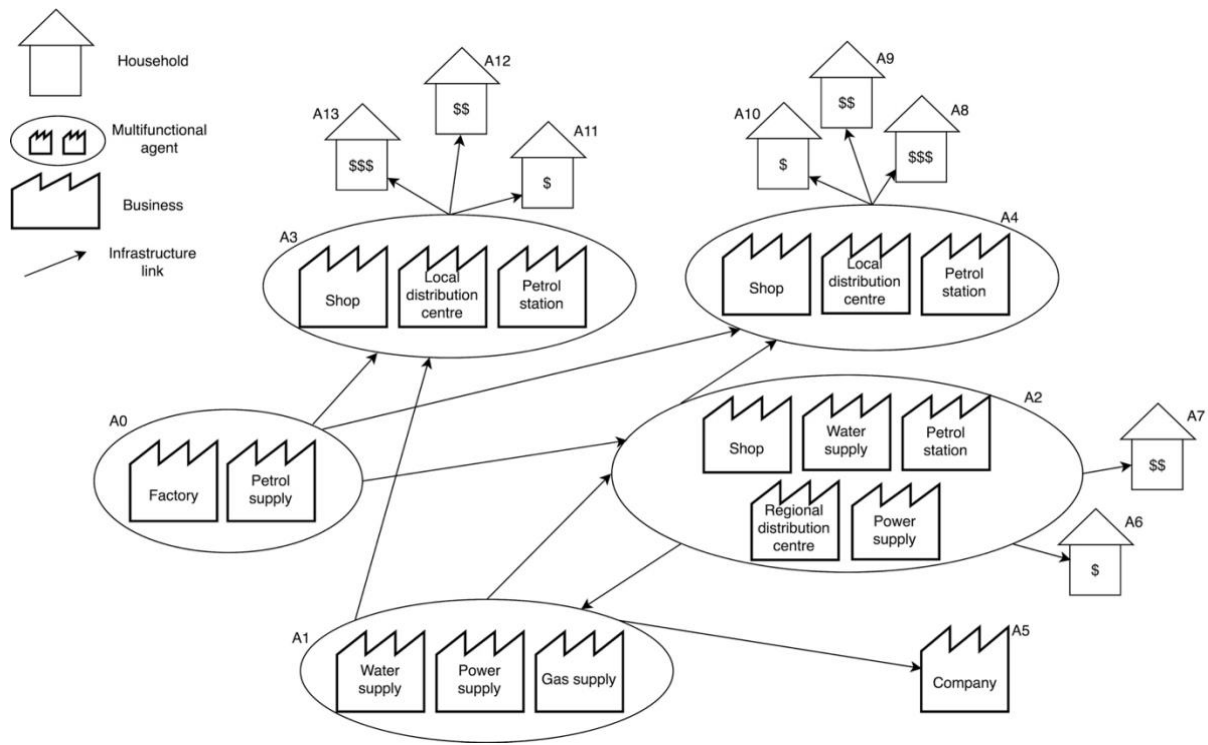

Figure 9: Case study of simulated system covering 6 types of resources (capital goods, consumer goods/services, water, power, gas, and petrol) transported over 16 infrastructure system links (roads, gas pipes, power lines, and water pipes) between 14 business and household agents. Different businesses at same location are grouped into aggregate multifunctional agents. Households are differentiated according to distinct demographic properties, here marked by $ signs signifying income levels.

In the network, two agents were attached to the providers of raw materials, thus allowing us to introduce new resources without undergoing another production process. Those resources were capital goods and petrol (agent A0); and water, gas, and electrical power (agent A1). Nine agents (A5-A13) were connected to the final consumer-demand vectors. They included eight for households and one for a business (i.e., company). All 14 agents were also permitted to produce certain resources as necessary. The rules for generating output resources, based on input resources, are listed in Table 1, as they applied to a single agent (A2), while the various specifications of roles for these agents are displayed in Table 2. These specifications are based on agents' roles as can be seen from FIGURE 9. Consequently, matrices of technical coefficients differed among agents. These matrices were devised to reflect a real-world concept that the production of most resources involves a relatively large quantity of one resource and smaller quantities of other resources (cf., Table 1). The actual numerical values are secondary to the outcome of this study, the relative magnitude of values is important and the change of these values, when a disruption is introduced. The matrices reflect production of resources as per FIGURE 9, where certain agents produce goods utilizing other goods, and certain agents just consume them. These follow the principle that most production utilizes primarily one resource with some additional inputs from other resources. Example of a matrix of technical coefficients constructed from FIGURE 9 following these principles is shown in Table 1. Some households were also able to produce certain goods as is normal due to domestic workers or household production. Each link in the system was assigned a vector to define its associated transportation cost and capacity. This was also done with the principle that commercial links have cheaper operating cost per unit than consumer links. Similarly, the actual numerical values are secondary for the purpose of this study, as their intention is to show relationships between agents and infrastructure links, and the relative impact of disruptions. In all, agents could use up to six resources in their production processes and could produce up to six types of resources.

|      |                              | To    |       |     |        |               |                                  |
|------|------------------------------|-------|-------|-----|--------|---------------|----------------------------------|
|      |                              | Power | Water | Gas | Petrol | Capital goods | Consumer goods and services      |
| From | Power                        | 0.18  | 0.90  | 0   | 0      | 0             | 0.20                             |
|      | Water                        | 0.30  | 0.10  | 0   | 0      | 0             | 0.30                             |
|      | Gas                          | 0.76  | 0.10  | 0   | 0      | 0             | 0.40                             |
|      | Petrol                       | 0.30  | 0.08  | 0   | 0      | 0             | 0.30                             |
|      | Capital goods                | 0.14  | 0.05  | 0   | 0      | 0             | 0.20                             |
|      | Consumer goods and services  | 0.10  | 0.05  | 0   | 0      | 0             | 0                                |

Table 1: Example matrix of technical coefficients of producing agent A2. Columns show amounts of resources needed to produce one unit of corresponding resource. In this case the agent is capable of producing power, water, and consumer goods and services.

|       |    | Resource |       |     |        |               |                             |
|-------|----|----------|-------|-----|--------|---------------|-----------------------------|
|       |    | Power    | Water | Gas | Petrol | Capital goods | Consumer goods and services |
| Agent | A0 |          |       |     | S      | S             |                             |
|       | A1 | SIO      | SI    | SI  | I      | I             | I                           |
|       | A2 | IO       | IO    | I   | I      | I             | IO                          |
|       | A3 | I        | I     | I   | I      | I             | O                           |
|       | A4 | I        | I     | I   | I      | I             | IO                          |
|       | A5 | IC       | IC    | IC  | IC     | IC            | OC                          |
|       | A6 | IC       | IC    | IC  | IC     | IC            | OC                          |
|       | A7 | IC       | IC    | IC  | IC     | IC            | OC                          |
|       | A8 | IC       | IC    | IC  | IC     | IC            | OC                          |
|       | A9 | IC       | IC    | IC  | IC     | IC            | OC                          |
|   | A10 | IC | IC | IC | IC | IC | OC |
|   | A11 | IC | IC | IC | IC | IC | OC |
|   | A12 | IC | IC | IC | IC | IC | OC |
|   | A13 | IC | IC | IC | IC | IC | OC |

Table 2: Specifications of system flows: S, supply of raw materials; I, input for production process; O, output from production process; and C, final consumer demand.

In choosing this particular topology as a prototype case study of our model, we were guided by principles meant to mimic a small system representative of a real-life geographical block in an urban area. As stipulated above (cf., FIGURE 9), only agents A0 and A1 were allowed to introduce raw materials into the system, while three business agents -- A2 through A4 -- performed the task of producing resources without the ultimate consumption of any goods. The largest group of agents (A5-A13) served as the final consumers, corresponding primarily to households or end-user businesses. The producing agents were connected with infrastructure system links to form a loop between them and to provide them with a choice of suppliers. This demonstrated the main functionalities and situations that the model might face in real-life scenarios, e.g., a wide selection of suppliers or the co-dependency found among businesses.

A system supply curve was derived by first setting the aggregate total demand and then measuring production costs aggregated across all resources for each consumer agent. This allowed us to shape the curve and determine the distribution of production costs across the final consumer agents for each scenario.

The above mechanism was implemented for various disruption scenarios to monitor possible changes in the supply curve and system performance. This was quantified with a metric to examine the cost of satisfying the system in response to a disruption. For purposes of comparison, we initially ran the system at normal operating capacity, without any disruptions. Final consumer demands were fixed and data were collected to estimate the cost of producing resources across all agents. After obtaining this default supply curve, we introduced eight different types of disruption, repeating the above process each time to acquire new supply curves for individual scenarios.

### Layout of the simulation experiment

The 3-by-3 factorial layout is shown in Table 3. For each scenario, the distribution of production costs served as our metric of performance. In developing our model, we predicted that the cost of producing the same quantity of resources would rise due to the increasingly negative impact of a disruption on the system. This layout was the smallest that could be applied for testing the features of our model prototype, i.e., confirming the effect of a disruption and revealing the emergent behavior of the system when multiple scenarios were combined. Its size was sufficient to understand the model and provide a convincing example of how it could be used by the scientific community and practitioners.

|  | | Production processes | | |
|---|---|---|---|---|
|  | | **Normal operation** | **Medium disruption** | **Heavy disruption** |
| **Infrastructure systems** | **Normal operation** | Base condition/No disruption | Scenario 2 | Scenario 6 |
|  | **Medium disruption** | Scenario 1 | Scenario 3 | Scenario 7 |
|  | **Heavy disruption** | Scenario 4 | Scenario 5 | Scenario 8 |

Table 3: 3x3-factorial layout for eight experimental scenarios of varying impacts compared with base conditions (normal operations). Scenarios 1 and 4: disruption to infrastructure system but not to production processes; Scenarios 2 and 6: disruption to production processes but not to infrastructure system; Scenarios 3, 5, 7, and 8: disruptions to both infrastructure and production processes.

### Disruption scenarios affecting infrastructure links

In Scenarios 1 and 4, we destroyed certain key connections between agents to simulate a disruption to infrastructure system links that would not inhibit the production processes themselves. These situations corresponded to, for example, a hurricane or a blizzard. In Scenario 1 (medium intensity), we introduced disruptions to two infrastructure system links between the agents (i.e., from A0 to A3, and from A0 to A4). This was done by setting the cost of transfer to infinity and the capacity to 0 for all resources. Hence, no resources could be moved through those links. This simulated a breakdown in infrastructure systems that prevented the passing of goods and services between households and businesses over a certain route that otherwise would be available. In Scenario 4 (heavy disruption), we additionally increased the costs for transporting all resources over an infrastructure link (from agent A2 to A1). This simulated a disruption in which only transfer costs were higher over a certain link.

### Disruption scenarios affecting production

In Scenarios 2 (medium disruption) and 6 (heavy disruption), we introduced system-wide changes that would only alter the amount of resources needed by the agents to produce other resources. Those disruptions were achieved by modifying the matrices of technical coefficients for agents A1 through A5. The scenarios corresponded to situations in which a household or business suffered a malfunction, such as a labor strike or equipment failure. Hence, the demands for inputs by the affected agents changed, along with the inputs into the system overall. Consequently, the agents' inputs increased because their processes required more resources to produce the same quantities of resources as before. Because its disruption was greater, Scenario 6 incurred larger changes in its matrices of technical coefficients.

### Disruption scenarios affecting infrastructure links and production

Scenarios 3, 5, 7, and 8 featured all combinations of concurrent disruptions to both the infrastructure system links and the agents. The matrices were modified for the agents while cutting the links so that production processes malfunctioned simultaneous to the breakdown of infrastructure systems. This mimicked a real-life, direct interruption that might affect the transportation network as well as some households and businesses, causing the movement of resources to be re-routed, such as due to an earthquake or a terrorist attack. Under such conditions, agent demands would likely also increase because production processes would have required more input resources to produce the same amount of output.

### Simulation results

We set the aggregate quantity of resources produced, represented by the aggregate final consumer-demand vectors (Table 4), to be the same for each scenario. Those selected vectors corresponded to consumers with different demand requirements. We wanted to achieve a situation where each final agent consumed at least a portion of each resource (some in large amounts, others in small quantities), thereby depicting a wide range of consumption patterns. Our primary goal was to exhibit a range of final consumer agents with different characteristics. Fixing the aggregate quantity of resources produced allowed us to obtain the cost distribution for supplying resources across agents in response to each type of disruption. Those costs shaped the supply curve of the system under each disruption scenario. However, the distribution of types of resources produced still varied among agents. The actual numerical values corresponding to demands are secondary to their relative differences and changes due to disruptions being introduced. We were primarily concerned with analyzing impacts of disruptions relative to other disruptions, and the response of the system to these.

|  | **A5** | **A6** | **A7** | **A8** | **A9** | **A10** | **A11** | **A12** | **A13** |
|---|---|---|---|---|---|---|---|---|---|
| **Power** | 9.75 | 10.75 | 9.75 | 9.75 | 9.75 | 9.75 | 9.75 | 9.75 | 9.75 |
| **Water** | 9.75 | 9.75 | 9.75 | 9.75 | 9.75 | 9.75 | 9.75 | 10.75 | 9.75 |
| **Gas** | 12.75 | 11.75 | 9.75 | 9.75 | 9.75 | 9.75 | 9.75 | 9.75 | 9.75 |
| **Petrol** | 13.75 | 9.75 | 9.75 | 9.75 | 9.75 | 9.75 | 9.75 | 9.75 | 9.75 |
| **Capital goods** | 17.75 | 9.75 | 9.75 | 9.75 | 9.75 | 9.75 | 9.75 | 9.75 | 10.75 |
| **Consumer goods and services** | 19.75 | 17.75 | 10.75 | 10.75 | 10.75 | 9.75 | 10.75 | 10.75 | 10.75 |

Table 4: Final consumer-demand vectors used in simulation experiments to derive supply curves for each scenario. Columns: vectors corresponding to agents (A) with final consumer demands; Rows, amounts demanded for respective resources.

Under all scenarios, the supply curve shifted upward, and its shape changed (FIGURE 10). This reflected the increased costs of producing the same level of resources when compared with normal operations. Whereas we had expected this upward shift in response to a disruption, we could not predict how the curve shape changed under our test scenarios. For example, the curve associated with normal operations was relatively flat but became steep as the production quantity rose. In contrast, the curve under Scenarios 1 and 4 (disruption only to infrastructure links) was flat and initially overlapped with the no-disruption-scenario curve. It then showed a larger increase before flattening out as the aggregate quantity increased. For Scenario 2 (medium disruption, only to production processes), the supply curve again shifted upward, signifying the rising cost to produce the same quantity of resources as under normal conditions. This shift was significantly larger than that revealed by Scenarios 1 and 4 but was still smaller than that found under other scenarios. The shape of the curve also changed drastically, being flat for low quantities but becoming steeper in the middle and again flattening toward the end. Similarly, in Scenario 6, the shift was significantly larger than in Scenario 2. This signified the strong impact that malfunctions had on production processes.

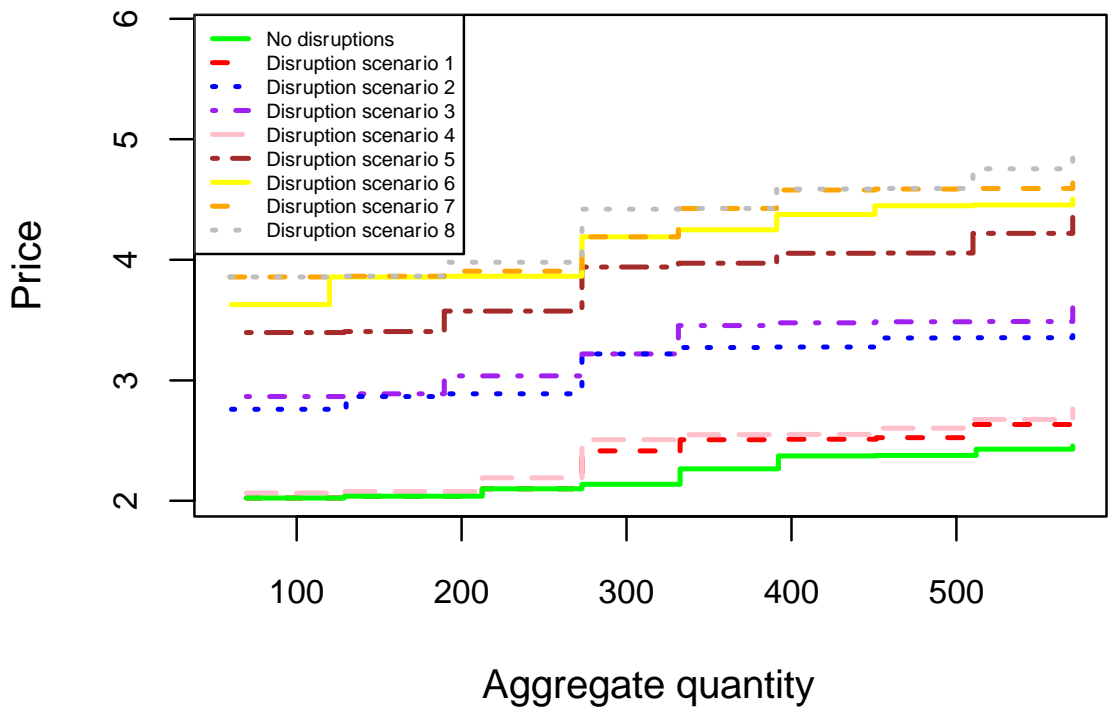

Figure 10: Cost of providing additional units of resources in response to system disruptions. Curves indicate prices based on given aggregate quantity of goods and services produced under various scenarios. Upward shifts in curves demonstrate higher costs associated with introduction of disruption.

The most dramatic change was associated with the introductions of Scenarios 7 and 8, in which both infrastructure and production processes were severely affected. When compared with normal operations, the significant upward shift in the supply curve corresponded to a major disruption in multiple parts of the system. This movement clearly indicated the severe impact these disruptions had on the system, which led to a rise in costs. The consequences of Scenarios 3, 5, 7, and 8 differed from those detected when only single impacts were combined. These findings demonstrated the significance of emergent behavior in our model, where failure to both infrastructure links and socio-economic units resulted in a disruption that was larger than the sum of disruptions to those components individually. The variations among curves in our model helped elucidate how an administrator might limit the quantity of a critical resource distributed to the population suffering a disruption so that one would not exceed the maximum cost that could be borne by the society. Those variations also could be used to assist in managing demand to ensure the delivery of a critical resource to the population.

We also applied other measures of performance in these experiments, assessing the total cost of satisfying the demand from the system based on a given configuration of final consumer demand from each particular node. The total cost of satisfying the system is the sum of costs across all agents. This entailed recording how a system would respond to a third type of disruption, that is, one that altered patterns of consumption. Table 5 presents the results when this metric was used for two different configurations under the eight disruption scenarios as well as during normal operations. The total cost of disruption was a sum of all the final demand vectors multiplied by the cost per unit of resource at the final consumer agent aggregated across all the agents and all the resources i.e. it was an aggregation of the final production cost of all resources across all agents.

| Aggregated final consumer demand from the system | Total cost under disruption scenario no.: | | | | | | | | |
|---|---|---|---|---|---|---|---|---|---|
| | None | 1 | 2 | 3 | 4 | 5 | 6 | 7 | 8 |
| 44 | 98.6 | 100.3 | 148.9 | 150.0 | 104.6 | 179.9 | 186.8 | 188.0 | 199.2 |
| 570.5 | 1274 | 1348 | 1798 | 1866 | 1383 | 2213 | 2378 | 2446 | 2491 |

Table 5: Total costs to satisfy aggregated final consumer demand from system -- normal conditions versus various disruption scenarios -- for two demand configurations.

Finally, performance was assessed as a percentage (fraction) of the demand that a system could provide under a particular disruption as compared with demand provided by an unstressed system at a given price. This represented the degradation of system performance from the basic scenario due to an introduced disruption. According to this metric, a price of 3.8 units was associated with 100% performance (i.e., 100% of demand satisfied) under normal, non-stressed conditions and also in Scenarios 1 through 4, versus only 33% performance under Scenario 5, 11% performance under Scenario 6, and 0% performance under Scenarios 7 and 8. This meant that, at a price of 3.8 units under Scenarios 7 and 8, the network completely ceased to function because it could not satisfy any of its demand at that price. Such a metric would appear to be useful in situations where the demand for resources is elastic and can be managed while supply is the main market force. Other available metrics might include calculating the percentage of the original price that the consumers would have to pay for the same quantity of resources after a disruption.

Analysis of FIGURE 10 revealed differently shaped supply curves for each scenario, an indication that those disruptions affected the system and agents in contrasting ways. We found this interesting because it suggested that the model could present not only general shifts in the curves but also changes in their shapes in response to system disruptions. This variability among curves was also closely related to the topology of the system. Hence, one limitation of our experiment was that it was performed under a single, defined topology. Although in scenarios where infrastructure links were disrupted, this topology was transformed into another by removing the edges, the base topology remained unchanged. Modifying the network topology and then performing the experiment with a completely different topology would likely have a significant impact on the shape and placement of those supply curves.

A simulation's runtime performance is affected by the size of the network and the number of resources included in the system. The problem is scalable because computation of each node is independent and can be distributed. Similarly, after each such computation, information is exchanged through network links. For larger networks, however, that exchange, or system synchronization, can require a significant amount of time. To accommodate that challenge due to size, households and businesses can be aggregated for a certain area or sector based on criteria of similarities or orthogonality between these units, e.g., location, income, health, or age (for households and locations), or the type of business (for company agents). Those aggregated, multifunctional agents are then worked into the network. Thus, scalability of the simulation is preserved at minimal expense to accuracy of the simulation under certain assumptions.

# Conclusions

The purpose of this study was to (1) develop an agent that mimicked the metabolism of a business or a household obtaining supplies from and providing output to infrastructure systems; (2) implement a network of agents that exchanged goods and services, as coordinated by a price mechanism; and (3) test the response of this prototype system to disruptions.

We achieved three main outcomes. First, we developed an agent that applied Leontief's input–output model to describe production as the transformation of supplied goods and services into output goods and services. The former included those supplied by infrastructure systems and by the production processes of other agents. The agent also interfaced with providers of raw materials and with final consumer demands.

Second, in developing our theoretical model, we set up a network of agents that utilized a price mechanism to coordinate the allocation of those goods and services when a self-organizing system was disrupted. Resources were dynamically assigned to agents based on production and transportation costs. This scalable network could be used with a varying number of resources and agents.

Third, we conducted simulation experiments to assess the feasibility of the model under normal operating conditions and also after introducing disruption scenarios that affected infrastructure systems, production processes within agents, or several combinations of the two. This allowed us to test our theoretical model under a simple abstract prototype application that corresponded to an urban setting. In order of impact, the situation in which both components (infrastructure links and socio-economic units) were influenced was the most damaging, followed by the scenario that interrupted only production processes.

In our new approach, we utilized Leontief's input–output model to represent a network of individual agents, and monitored their interactions when combined into a system-of-systems. Previous researchers tended to focus on modeling individual infrastructure systems and their resilience [55][56][57][11]. Others, such as Rinaldi et al. [4], described the interdependencies among infrastructure systems and showed how they could be thought of as flows. In contrast, Furuta et al. [58] portrayed interdependencies exhibited within a network of infrastructure systems as an SoS model. However, none of those studies included models of interactions between socio-economic units and infrastructure systems. This challenge was attempted by Colon et al. [32] in their proposed model that involved socioeconomic units, such as households and businesses, and a transportation network. They modeled supply chains within a country of Tanzania, by combining input-output model and network analysis.

However, they focused on only transportation system and did not include other infrastructure systems. In a similar fashion to our proposed framework they utilized input-output model and network analysis. However, their proposed model utilizes one large input-output model matrix to include all the companies present in the area considered and assuming that each company agent produces one resource. Hence, the spatial granularity of the model might not be easily adjusted and interchangeability of resources is limited. Moreover, contrary to our model, Colon et al.'s model allocates and prices resources based on suppliers' attempts to satisfy their predetermined profit margins rather than decrease the overall cost of producing the resources – a more appropriate approach for infrastructure providers. Colon et al.'s model includes stocks and inventories, households are limited to only consume resources and businesses to only produce. In contrast, under our model both these activities could be achieved by any agent, and we do not provide a mechanism for stocks of resources as these are often impracticable for infrastructure resources. Finally, the model presented by Colon et al. is derived chiefly from economic data rather than stemming primarily from physical and engineered infrastructures as its backbone. Therefore, our proposed strategy is unique in that it combined a model of household and business agents with a model for self-organizing distribution networks of multiple infrastructure systems between those agents under disruptions. In doing so, our model could differentiate between population groups and types of businesses, a feature omitted in regional-level models.

Our self-organizing adaptation of the model to disruptions, as presented here, is novel. Cascading failures that are propagated through several infrastructure systems and their components have already been examined [e.g., see Dueñas-Osorio and Vemuru [9]]. Kotzanikolaou, Theoharidou, and Gritzalis [59] have analyzed the interdependencies among infrastructure systems to measure the impact that such cascading failures might have. Likewise, Rinaldi and colleagues [4][22] have investigated interdependent infrastructures in an effort to improve their understanding about failures and their influence on those systems. These topics are of great concern to governments, businesses, and the public [60] and we have now expanded upon the results from those earlier studies to develop a self-organizing mechanism for adaptations to disruptions.

Since its first mention by Leontief [61], the input–output model has been applied to describe the production of resources by geographical area in economies [62]. It has also been utilized to investigate the inoperability of interdependent economic sectors due to infrastructure disruptions [29][63][32]. In the field of supply-chain management, this model has been used to determine different production processes within a business [38][49][64]. By implementing the input-output model with agents to simulate those processes by a business or a household interacting with infrastructure systems, we have confirmed that Leontief's model is effective when extended to the interfaces that an agent might have with suppliers of raw materials, final consumer demands, and other agents.

The results of our work have several implications for scientists, policy-makers, infrastructure designers, and businesses. For scientific endeavors, the model presents a scalable approach to evaluating infrastructure systems, their interactions with households and businesses, and the impact(s) that disruptions have on systems at a household/business level of granularity. Members of the scientific community could use the model to determine how different demographics are affected by those disruptions. Policy-makers can employ this as a support tool to aid in making crucial decisions about infrastructure development, real-time assessment of the impact of disruptions, or the creation of contingency plans against future disruptions. Including households and businesses in the analysis allows planners to understand the population groups that are most affected by different types of disruptions because our model can illustrate the impact that disruptions have on the most vulnerable groups within a society. Decision-makers will also gain a greater understanding of how disruptions affect a company, based on regional characteristics as well as on its specific type of business.

For practitioners, this tool can be applied for stress-testing and assessing their systems, or for designers who can compare topologies and different components within a system. The model can assist in examining different topologies, introducing redundancy or more capacity/lower price to various components of the network, and then estimating the impact of these under disruptions. This in turn can help in making decisions as to where a system needs reinforcement or available infrastructure investment resources should be allocated. The costliest vulnerabilities to a system can be identified according to established criteria, based on data collected from an analysis of that system when challenged with a wide range of randomly generated disruptions. Performing such experiments enables planners to predict the value at risk for the system as well as the potential costs associated with the most devastating vulnerabilities. Ultimately, the results of a simulation implementing our model might be used in determining, for example, where to build a bridge or how to expand a power grid.

The importance of demand management has been noted in the literature [e.g., see [65][66]]. Our model can help users understand the response of a system to disruptions under different demand conditions and detect the impact that managing demand has on the cost of satisfying the system. The model can help to manage demand to ensure the delivery of critical resources and services under disruption, when resources are limited. Finally, this model can be applied to studying the effects of business discontinuities, such as those that arise due to shortages or an increase in the price of resources. Those findings can then be incorporated into business contingency planning.

The primary limitation of our experiment was the use of only one topology for testing, which may have influenced model performance. Different infrastructure systems have different topologies that vary with each other, hence duality with the real world was not necessarily fully preserved. Furthermore, disruptions in real-life scenarios can vary widely, and our selection of only eight scenarios and disruption logic could not cover all possible cases of system failures, which can vary and be domain-specific. In addition, resource allocation might not always follow the lowest cost source. Our model did not consider regulatory frameworks or other social factors that can affect the delivery of resources. However, the theoretical model could take these factors into account through different cost functions. In real-world settings, production can, of course, be more complex and involve non-linear factors that cannot be easily described with an input–output model. However, linear models presented in this study present sufficient approximations of production processes. The transportation and supply of raw materials, as well as final consumer demands, can also be non-linear, making them more complex than the linear models and scalars that we employed when evaluating our prototype. However, the approximations presented in the small-scale prototype are appropriate for such models. Moreover, the recovery planning process and recovery capabilities of damaged components were not considered in our model. The self-organizing behavior of our model could represent recovery of critical functions under disruptions. However, the planning for recovery such as adjusting individual components of the system in response to a disruption or developing certain disruption response policies and strategies was outside of the scope of this study.

Several challenges were identified here that require further examination. Proposed projects might involve extending experiments to evaluate the model under additional topologies to investigate the impact of network topology on the systems. Different topologies can alter dispersion of resources throughout the system, thus influencing the cost of delivery of resources to nodes of the network. This would result in widely different result of the simulation experiments. Furthermore, we could expand upon the mechanism for resource allocation. In particular, for emergency scenarios, mechanisms other than cost are often employed for such allocations. Future research could also consider a wider range of disruption logic and scenarios that provide better coverage of actual situations. More mechanisms could be included for defining raw material inputs and final consumer demands in the system. This would also correspond more closely to real-world challenges. Finally, the recovery planning and disruption response policies could be tested and included within our model.

## Acknowledgement
This work is an outcome of the Future Resilient Systems project at the Singapore-ETH Centre (SEC), which is funded by the National Research Foundation of Singapore (NRF) under its Campus for Research Excellence and Technological Enterprise (CREATE) programme (FI 370074011). We would like to thank Priscilla Licht for help in proof-reading and language editing.